\documentclass[pra,aps,showpacs,floatfix,nofootinbib,twocolumn,superscriptaddress]{revtex4-1}
\usepackage{amsmath}
\usepackage{amsfonts}
\usepackage{amssymb}
\usepackage{revsymb}
\usepackage{graphicx}
\usepackage{mhchem}
\usepackage{siunitx}
\usepackage{bm}
\usepackage{dcolumn}
\usepackage{multirow}

\begin{document}
\title{An extrapolation method for polarizability assessments of ion-based optical clocks}
\author{M. D. Barrett}
\email{phybmd@nus.edu.sg}
\affiliation{Centre for Quantum Technologies, National University of Singapore, 3 Science Drive 2, 117543 Singapore}
\affiliation{Department of Physics, National University of Singapore, 2 Science Drive 3, 117551 Singapore}
\author{K. J. Arnold}
\affiliation{Centre for Quantum Technologies, National University of Singapore, 3 Science Drive 2, 117543 Singapore}
\begin{abstract}
We present a numerical method for extrapolating polarizability measurements to dc as done in the assessment of blackbody radiation shifts for ion-based clocks.  The method explicitly accounts for the frequency dependence of relevant atomic transitions without introducing an ad hoc modelling function.  It incorporates \emph{a priori} atomic structure calculations, which allows measurements to be augmented by calculations if there is insufficient data to make a purely measurement based estimate.  The method also provides indicators of inconsistencies between theory and experiment or inadequacies of the data for making an extrapolation.  We use results from Al$^+$, Lu$^+$, and Yb$^+$ to illustrate features of the method.  
\end{abstract}
\maketitle
The blackbody radiation (BBR) shift is an important systematic in most optical atomic clocks.  It typically requires accurate determination of the differential scalar polarizability, $\Delta\alpha_0(\omega)$, which characterizes the sensitivity of the atomic clock transition to the thermal radiation field.  For ion-based clocks, one approach has involved measurements at near infra-red (NIR) frequencies and subsequent extrapolation to dc, with the extrapolation involving the introduction of a model of $\Delta\alpha_0(\omega)$ over the range of interest \cite{arnold2019dynamic,huntemann2016single}.  In the case of Al$^+$, a single measurement point was used to constrain theory and the possible variations of $\Delta\alpha_0(0)$ that the measurement imposed \cite{rosenband2006blackbody}.

When extrapolating from a single data point, one must be reliant on theory to some degree.  If the measurement point is far removed from all contributing atomic transitions, then physically the dc value cannot be so far from the measured value.  If theory is consistent with the measurement or accurately predicts related atomic properties, it would be reasonable to consider what small corrections theory might allow when constrained by the measurement.  When introducing a model to fit the data, some level of justification for the model must also be based on a theoretical description of the atom, or at least account for physical constraints based on the atomic structure.

Here we consider the question in more general terms: given a set of measurements of $\Delta\alpha_0(\omega)$ and an \emph{a priori} set of estimates of the contributions to $\Delta\alpha_0(\omega)$, what is the best estimate that can be made for $\Delta\alpha_0(0)$.  The method we use considers the measurements as projections onto to an over-complete set of basis functions defined by the allowed atomic transitions and progressively eliminates the dependence on theory as more measurement points are added.  Consequently, it does not require accuracy of the calculations, provided there are sufficient measurements spanning a suitable measurement window, as needed for any extrapolation.  Inclusion of the available calculations provides indicators of inconsistency between theory and experiment or inadequacies of the experimental data, which we illustrate by example.  Alternatively, calculations can be used to estimate a correction if there is insufficient data to make an accurate extrapolation as done for Al$^+$ \cite{rosenband2006blackbody}.
\section{The projection method}
In general, $\Delta\alpha_0(\omega)$ is given by 
\begin{equation}
\label{PolEq}
\Delta\alpha_0(\omega)=\sum_m\frac{c_m}{1-(\omega/\omega_{0,m})^2}
\end{equation}
where $\omega_{0,m}$ and $c_m$ are, respectively, the resonances and weights of all contributing transitions.  In principle the summation should include an integral to account for transitions to the continuum.  To a good approximation, these terms, along with other core excitation and correction terms, can be treated as constants with the associated resonant frequencies taken to infinity.  As written, contributions from the lower state have $c_m<0$, although our method does not enforce this condition.  

It is convenient to express Eq.~\ref{PolEq} in vector form
\begin{equation}
\Delta\alpha_0(\omega)=\sum_m\frac{c_m}{1-(\omega/\omega_{0,m})^2}=\mathbf{f}(\omega)\cdot \mathbf{c},
\end{equation}
where the $m^{th}$ component of $\mathbf{f}$ is $1/(1-(\omega/\omega_{0,m})^2)$.  Since transition frequencies, $\omega_{0,m}$, are generally well-known, $\mathbf{f}(\omega)$ is a practically exact set of basis functions.  A given measurement  $m_j=\Delta\alpha_0(\omega_j)$ can then be viewed as a projection of $\mathbf{c}$ onto $\mathbf{f}(\omega_j)$.  An extrapolation to dc then amounts to determining the projection of $\mathbf{c}$ onto $\mathbf{f}(0)$  from a given a set of projections $m_j=\mathbf{c}\cdot\mathbf{f}(\omega_j)$ each with uncertainty $\sigma_j$.

The basic idea of our approach is to make use of the fact that $\mathbf{f}(0)$ can be written as a sum of two orthogonal vectors, $\mathbf{u}$ and $\bar{\mathbf{u}}$, where $\mathbf{u}$ is in the vector space spanned by the set of all $\mathbf{f}(\omega_j)$ and $\bar{\mathbf{u}}$ within the complement.  An estimate for $\Delta\alpha_0(0)$ can then be found from
\begin{equation}
\label{decomp}
\Delta\alpha_0(0)=\mathbf{c}\cdot\mathbf{f}(0)=\mathbf{c}\cdot\mathbf{u}+\mathbf{c}\cdot\bar{\mathbf{u}},
\end{equation}
with the first term estimated from measurements $m_j$ and the second from theory.  Provided there are sufficiently many measurements over a sufficiently large measurement window, the residual term $\mathbf{c}\cdot\bar{\mathbf{u}}$ would be zero or at least sufficiently small that it can be adequately bounded from theoretical considerations or neglected relative to the uncertainty in evaluating the first term.

The decomposition in Eq.~\ref{decomp} is best found by introducing the matrix $\mathbf{F}$ given by
\begin{align}
\mathbf{F}&=\left[\frac{\mathbf{f}(\omega_1)}{\sigma_1} \cdots \frac{\mathbf{f}(\omega_n)}{\sigma_n}\right],
\end{align}
and its singular-value-decomposition (SVD) $\mathbf{F}=\mathbf{U}\mathbf{W}\mathbf{V}^T$.  Properties of a SVD provide a construction for $\mathbf{u}$, namely
\begin{equation}
\label{range}
\mathbf{u}=\sum_{j=1}^k \left[\mathbf{u}_j \cdot \mathbf{f}(0)\right]\mathbf{u}_j,
\end{equation}
where $\mathbf{u}_j$ is the $j^\mathrm{th}$ column of $\mathbf{U}$ and $k$ is the number of non-zero singular values.  Solving $\mathbf{F}\cdot\mathbf{x}=\mathbf{u}$ gives
\begin{equation}
\label{sol}
\mathbf{x}=\sum_{j=1}^k \frac{\mathbf{u}_j\cdot \mathbf{f}(0)}{w_j} \mathbf{v}_j,
\end{equation}
where $\mathbf{v}_j$ is the $j^\mathrm{th}$ column of $\mathbf{V}$, and $w_j$ is the corresponding non-zero singular value.  We then have
\begin{equation}
\label{estimate}
\Delta\alpha_0(0)=\mathbf{\bar{m}}\cdot\mathbf{x}+\mathbf{c}\cdot\bar{\mathbf{u}},
\end{equation}
where the $j^\mathrm{th}$ element of $\mathbf{\bar{m}}$ is $m_j/\sigma_j$.  

Assuming the measurements are uncorrelated, the uncertainty in the first term is $\|\mathbf{x}\|$ and use of a SVD provides an optimal weighting of measurements $m_j$ to minimize this uncertainty.  Small singular values, which can degrade the estimate, may be omitted from Eq.~\ref{range} and \ref{sol}, provided the residual term $\mathbf{c}\cdot\bar{\mathbf{u}}$ remains negligible or can be determined from theory.  As expected for a linear combination of basis functions, terms in Eq.~\ref{sol} do not depend on the measurements themselves; they depend only on measurement uncertainties and the frequencies at which the measurements are made.  Moreover, each term in Eq.~\ref{sol} is orthogonal so the uncertainty contributions arising from each are independent.  We can expect that the estimate $\bar{\mathbf{m}}\cdot\mathbf{x}$ be dominated by the first few terms with additional terms of increasing $k$ providing corrections of decreasing statistical significance.  Ideally, $k$ would be chosen such that its contribution to the estimate is statistically significant and the remaining $\mathbf{c}\cdot\bar{\mathbf{u}}$ negligibly small.

By construction, the residual term in Eq.~\ref{estimate} is independent of the first so that uncertainties from the two terms in Eq.~\ref{estimate} can also be added in quadrature.  However, we are more concerned with the case in which the residual term is negligible so that it does not contribute to the estimate. In assessing the residual term it is important to note that a small value may well be the result of a fortuitous cancelation of terms.  Thus, it is important to consider reasonable variations in the theoretical estimates of $\mathbf{c}$.  As we are primarily interested in bounding the residual term, we consider two figures of merit,
\begin{equation}
\sigma_\mathrm{rms} =\sqrt{\mathbf{c}^2\cdot\bar{\mathbf{u}}^2},\quad\mbox{and} \quad \sigma_c=|\mathbf{c}|\cdot|\bar{\mathbf{u}}|,
\end{equation}
where the square and absolute operations are to be taken element-wise.  If all elements $\mathbf{c}$ are known with a fractional inaccuracy $\beta$, then $\beta\sigma_\mathrm{rms}$ is the uncertainty in the residual term assuming all estimates of $\mathbf{c}$ are uncorrelated.  Similarly $\beta\sigma_c$ would be the uncertainty in the residual term assuming a worst case correlation in the uncertainties of $\mathbf{c}$.

\section{{Al}$^+$}
The simplest system for which to apply the formalism is that of Al$^+$.  In this case, all transitions are deep in the ultra-violet (uv) so that, in the NIR, $\Delta\alpha_0(\omega)$ can be well approximated by an even order quadratic.  The formalism presented here can then be checked against that obtained via a quadratic fit. We consider measurements made at 1560 and 780\,nm with mean values given by theoretically calculated values and a measurement uncertainty of 10\%.  These wavelengths are chosen as they are readily available and span a reasonable measurement window.

To apply the formalism, we use calculations from \cite{safronova2012blackbody}.  Each of the contributions labelled ``other'' are treated as a single transition with a frequency given by the largest available wavelength for that contribution, and valence-core corrections are treated as a single constant giving eight basis functions in total.  Applying the projection formalism gives a mean estimate for $\bar{\mathbf{m}}\cdot\mathbf{x}$ of $0.494\,\mathrm{a.u.}$ with $\|x\|=0.072\,\mathrm{a.u.}$, where $\mathrm{a.u.}$ denotes atomic units.  The residual term, $\sigma_\mathrm{rms}$, and $\sigma_c$ are found to be $0.001, 0.001,$ and $0.0015$, respectively, so the residual term can be safely neglected.  The estimate is consistent with the theoretical value of $\Delta\alpha_0(0)=0.495\,\mathrm{a.u.}$, and the uncertainty is consistent with that expected for a quadratic fit, which gives a mean value of $0.492\,\mathrm{a.u.}$ with an uncertainty of $0.073\,\mathrm{a.u}$.  The slight bias in the estimate from a quadratic fit arises from the increasing importance of quartic terms at 780\,nm.  The projection formalism better handles this as it makes no such quadratic approximation.  This effect becomes more prominent as the measurement window is extended to smaller wavelengths.

In the case of Al$^+$, an alternative formalism has been used to extrapolate from a single measurement \cite{rosenband2006blackbody,brewer2019quantum}.  The formalism here also handles this case, which results in an expression
\begin{equation}
\label{single}
\Delta\alpha_0(0)=a_1\Delta\alpha_0(\omega_1)+\mathbf{c}\cdot\bar{\mathbf{u}},
\end{equation}
where $\Delta\alpha_0(\omega_1)$ is the measured value at $\omega_1$.  This equation has the same form as \cite[Eq.~S7]{brewer2019quantum} but differs by the manner in which $a_1$ is constructed. The derivation of \cite[Eq.~S7]{brewer2019quantum} is exact and holds for any value of their parameter $\delta_0=(1-a_1)/a_1$, which is chosen to minimize error contributions from the strongest transitions as noted in \cite{rosenband2006blackbody}.  This leads to an apparent ambiguity between the two approaches: here $a_1$ is independent of any uncertainties in the theory, whereas $\delta_0$ in \cite{rosenband2006blackbody,brewer2019quantum} is explicitly chosen based on those uncertainties.  Resolution of this ambiguity lies in the error analysis.  When there is only one measurement, there is one and only one choice of $a_1$ for which the two terms in Eq.~\ref{single} are uncorrelated and that is as constructed here.  For any other choice, the uncertainties from each term cannot be added in quadrature as done in \cite[Eq.~S8]{brewer2019quantum}.  As the value of $a_1$ for their measurement wavelength is close to 1 anyway, proper accounting for the correlation is unlikely to significantly affect their uncertainty assessment.  However, their formalism should not be used on general principle, as it is both unnecessary and incorrect.
\section{Lu$^+$}
A more comprehensive test of the formalism can be done using the results obtained with Lu$^+$ \cite{arnold2018blackbody,arnold2019dynamic}.  In this case measurements have been carried out over a large measurement window, and individual matrix elements corresponding to the dominant NIR frequency dependence of $\Delta\alpha_0(\omega)$ have been measured directly.   Extrapolation based on two different models gave consistent results \cite{arnold2019dynamic} with both models justified theoretically.  In addition there has been a number of results demonstrating consistency between experiment and theory.  

To apply the formalism, we use the theoretical results given in \cite{paez2016atomic}.   Contributions labelled as ``other'' are treated as a single constant, noting that they have transition wavelengths below $200\,\mathrm{nm}$ and there are 15 other basis functions spanning a wide wavelength range.  Measurements are taken from \cite{arnold2018blackbody,arnold2019dynamic}. The determination of reduced matrix elements are included by considering them measurements of the polarizability contribution from the corresponding transition with $\mathbf{f}(\omega_j)$ given by the unit vector having a 1 at position $j$.  We include also the reduced matrix element $\langle{}^3P_1\|r\|{}^1S_0\rangle$, which may be inferred from results in \cite{arnold2019dynamic} along with branching ratios reported in \cite{paez2016atomic}, giving a total of eight measurements.

Applying the formalism, we find eight non-zero singular values spanning almost eight orders of magnitude and it is instructive to tabulate results for increasing values of $k$.  These results are tabulated in table~\ref{LuResults} where we give $\bar{\mathbf{m}}\cdot\mathbf{x}$, $\|\mathbf{x}\|$, $\mathbf{c}\cdot\bar{\mathbf{u}}$, $\sigma_\mathrm{rms}$, $\sigma_c$, and $\Delta\alpha_0(0)$ when including $k=2,\ldots,8$ singular values. Independent of $k$, estimates are consistent and the $k=5,$ and $6$ results are completely consistent with the values reported in \cite{paez2016atomic}.  This is not surprising given the agreement between theory and experiment and the fact that the measurement window extends to near dc ($\lambda=10.6\,\mathrm{\mu m}$).  At $k=5$, the dependence on theory is almost completely eliminated with $\sigma_c\lesssim \|\mathbf{x}\|$.  Given that there has been a fairly consistent agreement between theory and experiment, $\sigma_c$ should be seen as an overly conservative estimate of the error.  This is supported by the consistent result from $k=6$, which practically eliminates any dependence on theory even though the correction has no statistical significance.  Subsequent terms clearly degrade the estimate, but still maintain consistency with the $k=5$ result.

As discussed in \cite{arnold2019dynamic}, the frequency dependence of $\alpha_0(\omega)$ is predominately determined by contributions from two poles at $598$ and $646\,\mathrm{nm}$.  All other contributions can be modelled by an additional quartic polynomial.  As theory predicted, there is a high degree of cancellation of the quadratic term of this polynomial, such that the measurement at $\lambda=10.6\,\mathrm{\mu m}$ is essential for the extrapolation.  Consequently it is instructive to consider the impact of this term in the framework of the projection method.  In table~\ref{LuResults2} we give results when omitting this measurement value for the most relevant values of $k$.  The $k=5$ term reduces the theoretical contribution to a value that is likely not significant given the value of $\|\mathbf{x}\|$ and $\sigma_c$. The estimate is at least consistent with the value reported in \cite{arnold2019dynamic} albeit with a larger uncertainty that leaves the sign of $\Delta\alpha_0(\omega)$ indeterminate.  However, one would need to justify the uncertainty given to the theory term.  Alternatively one could use the $k=6$ term, although it has no statistical significance.  Objectively, it should be seen for what it is: there is insufficient data to make a purely measurement-based estimate with a useful uncertainty.

\begin{table}
\caption{Extrapolations of $\Delta \alpha_0(\omega)$ to dc using the projection method on Lu$^+$ measurements.  Measurements are taken from \cite{arnold2018blackbody,arnold2019dynamic}, including the reduced matrix element $\langle{}^3P_1\|r\|{}^1S_0\rangle$, which may be inferred from results in \cite{arnold2019dynamic} along with branching ratios reported in \cite{paez2016atomic}.}
\begin{ruledtabular}
\begin{tabular}{c c c c c c c}
$k$ & $\bar{\mathbf{m}}\cdot\mathbf{x}$ & $\|\mathbf{x}\|$ & $\mathbf{c}\cdot\bar{\mathbf{u}}$ & $\sigma_\mathrm{rms}$ & $\sigma_c$ & $\Delta\alpha_0(0)$\\ [0.1pc]
\hline \\ [-0.6pc]
2 & 0.0503 & 0.0040 & -0.0319 & 0.0298 & 0.0653 & 0.0184 \\
3 & 0.0400 & 0.0040 & -0.0213 & 0.0283 & 0.0657 & 0.0187 \\
4 & 0.0312 & 0.0040 & -0.0116 & 0.0082 & 0.0203 & 0.0195 \\
5 & 0.0204 & 0.0041 & -0.0007 & 0.0015 & 0.0035 & 0.0197 \\
6 & 0.0202 & 0.0044 & - & - & 0.0001 & 0.0203 \\
7 & 0.0158 & 0.0052 & - & - & - & 0.0158 \\
8 & 0.0129 & 0.0152 & - & - & - & 0.0129 \\
\end{tabular}
\end{ruledtabular}
\label{LuResults}
\end{table}

\begin{table}
\caption{Extrapolations of $\Delta \alpha_0(\omega)$ to dc using the projection method on Lu$^+$ measurements.  Measurements are as for table~\ref{LuResults} but omit that at $10.6\,\mathrm{\mu m}$.}
\begin{ruledtabular}
\begin{tabular}{c c c c c c c}
$k$ & $\bar{\mathbf{m}}\cdot\mathbf{x}$ & $\|\mathbf{x}\|$ & $\mathbf{c}\cdot\bar{\mathbf{u}}$ & $\sigma_\mathrm{rms}$ & $\sigma_c$ & $\Delta\alpha_0(0)$\\ [0.1pc]
\hline \\ [-0.6pc]
4 & 0.641 & 0.053 & -0.624 & 0.467 & 1.083 & 0.017 \\
5 & 0.087 & 0.093 & -0.047 & 0.082 & 0.200 & 0.040 \\
6 & 0.464 & 0.276 & 0.009 & 0.006 & 0.011 & 0.473 \\
\end{tabular}
\end{ruledtabular}
\label{LuResults2}
\end{table}

\section{Yb$^+$}
A more challenging example is given by Yb$^+$, in which it is known that theoretical estimates of $\Delta\alpha_0(\omega)$ are in disagreement with theory and there is a small number of measurements.  To apply the formalism, we take theoretical results from \cite{biemont1998lifetime}, which reproduces the dashed curve given in \cite[Fig.~2]{huntemann2016single}.  Measurements are read from the latter figure with a best-effort resolution and are given in table~\ref{YbData} for reference purposes.  In order to make a fair comparison of the methodologies, it is necessary to elaborate on two short-comings of the assessment in \cite{huntemann2016single}.  The first concerns the statistical validity of the extrapolation, and the second concerns the validity of the single pole model (SPM),
\begin{equation}
\Delta\alpha_0(\omega)=c_0+c_1\frac{(\omega/\omega_0)^2}{1-(\omega/\omega_0)^2}.
\end{equation}

\begin{table}
\caption{Yb$^+$ measurement data read from \cite[Fig.~2]{huntemann2016single}.  Uncertainties are as stated in \cite{huntemann2016single}.}
\begin{ruledtabular}
\begin{tabular}{l c c c c}
$\lambda$(nm) & 1554 & 1310 & 1064 & 852 \\ [0.1pc]
\hline \\ [-0.6pc]
$\Delta\alpha(\omega)\,\mathrm{(a.u.)}$  & 4.899 & 4.786 & 4.278 & 3.315 \\
Error (\%) & 3 & 0.5 & 3 & 3 \\
\end{tabular}
\end{ruledtabular}
\label{YbData}
\end{table}

Fitting data in table~\ref{YbData} to the SPM gives an estimate $\Delta\alpha_0(0)=c_0=5.43(30)\,\mathrm{a.u.}$, where the uncertainty is derived using standard statistical methods and verified by Monte-Carlo simulation.  The estimate is within $1\%$ of the value $5.385(97)\,\mathrm{a.u.}$ reported in \cite{huntemann2016single} but the uncertainty is approximately three times larger.  Note that the estimate is based on a three parameter fit to four data points. Taking the result with $0.5\%$ inaccuracy as exact, we can eliminate a parameter leaving three data points with 3\% inaccuracy and a two parameter fitting function.  Consequently, an inaccuracy of $3\%$ for the extrapolation would be questionable moreso the $1.8\%$ reported in \cite{huntemann2016single}.

\begin{figure*}
\begin{center}
  \includegraphics[width=0.32\textwidth]{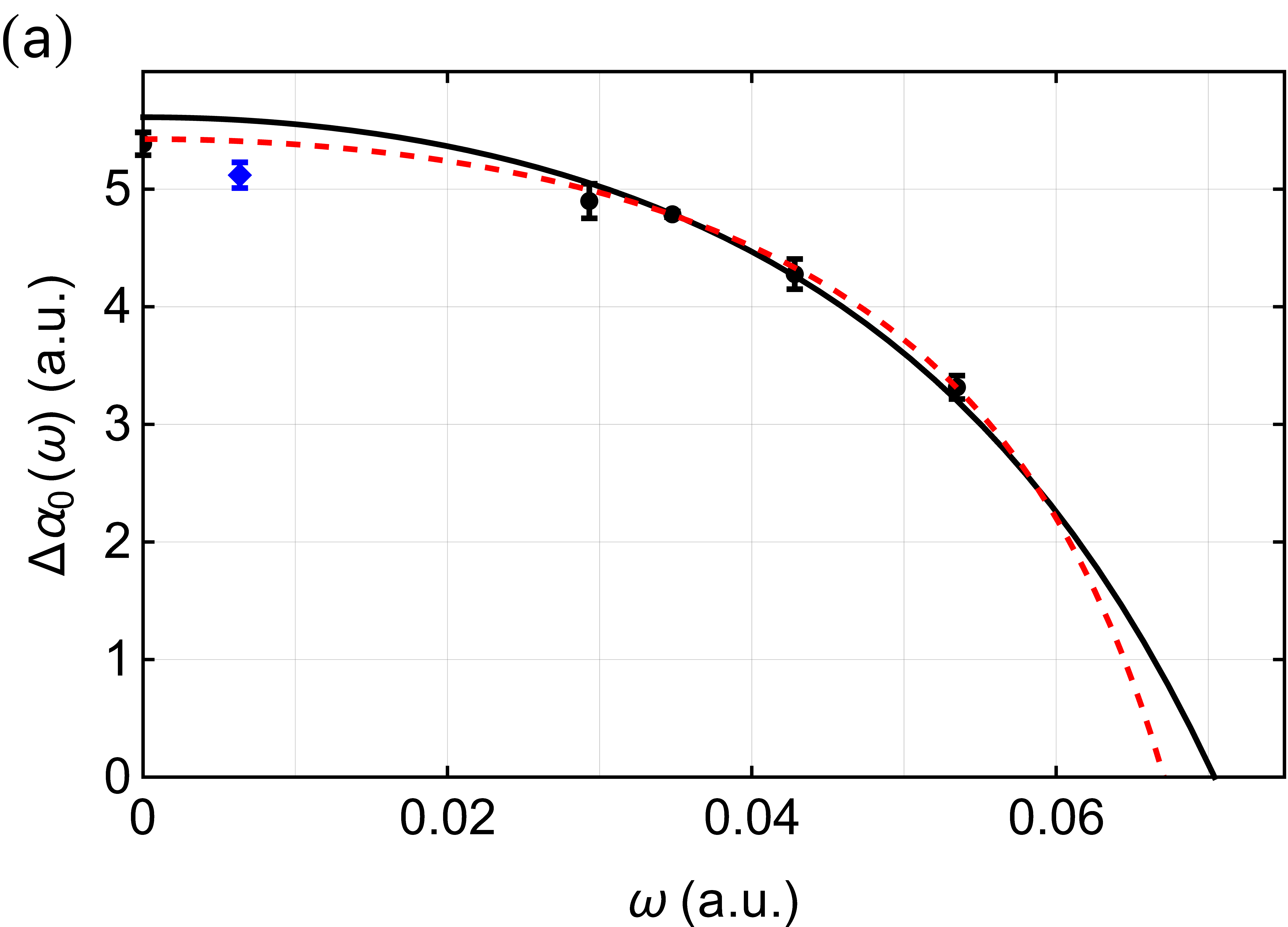}
  \includegraphics[width=0.32\textwidth]{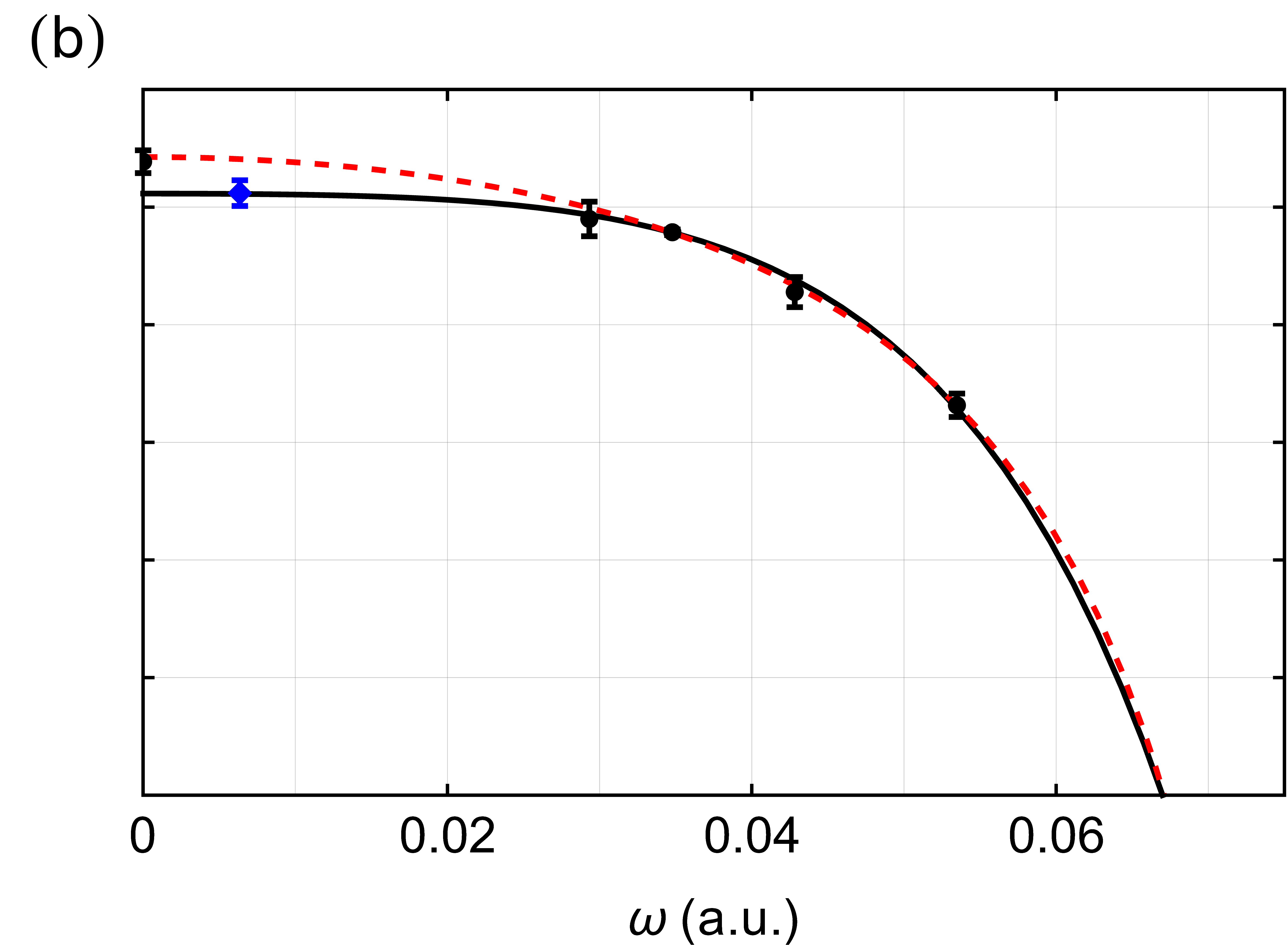}
  \includegraphics[width=0.32\textwidth]{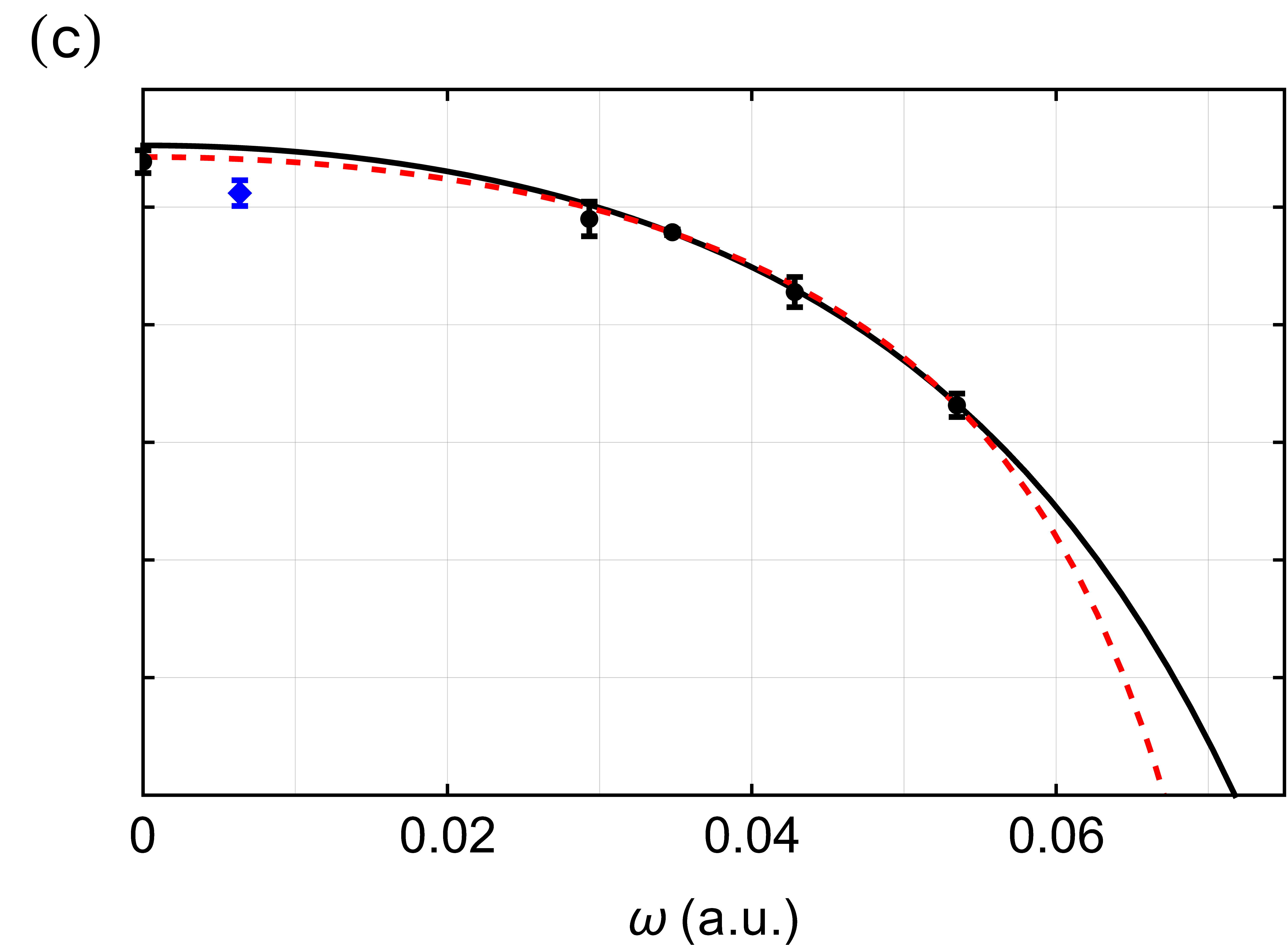}
  \caption{Polarizability fits for Yb$^+$.  In each plot the black dots are taken from table~\ref{YbData} in addition to the extrapolated point reported in \cite{huntemann2016single}, the blue diamond is value reported in \cite{baynham2018measurement}, and the red dashed curve is the fit to a SPM as used in \cite{huntemann2016single}.  The black curves are fits using: (a) the calculated polarizability with a variable offset used as a fitting parameter, (b) a DPM with a zero-crossing at 680\,nm giving $c_f=222$, and $c_g=150$, and (c) a DPM with a zero-crossing at 635\,nm giving $c_f=59$, and $c_g=49$.  Fits to the DPM used fixed values of the effective poles as discussed in the text.}
 \label{offset}
\end{center}
\end{figure*}

One should also carefully consider the validity of the SPM.  In \cite{huntemann2016single}, the authors justify the SPM by using it to fit data calculated from theory, noting that it gives an extrapolated value within 0.2\% of the calculated value and assume this as a modeling uncertainty.  In Fig.~\ref{offset}(a), we plot the data in table~\ref{YbData} along with a fit consisting of the calculated $\Delta\alpha_0(\omega)$ with a variable offset, that is, we assume the theory is correct up to an offset.  A $\chi^2$-fit gives an offset of $5.828(23)\,\mathrm{a.u.}$ with a corresponding dc extrapolation of $5.612(23)\,\mathrm{a.u.}$ and a $\chi^2$-statistic of 2.38 for three degrees-of-freedom (dof).  The probability of a $\chi^2$-statistic with 3 dof being above or below 2.38 is 50\% so the fit is statistically reasonable and the extrapolated value no less credible than any other.  However, we would not suggest that the model is perfect or that the derived uncertainty for the extrapolated value, which is almost completely determined by the one measurement, is meaningful.  This highlights the well-known fact that, just because a model fits to set of data, it does not mean the model is accurate or even correct.

The SPM can be viewed as a Pad\`e approximant of $\Delta\alpha_0(\omega)$ valid to 4th order, with the constraint that the quadratic and quartic terms have the same sign.  Given the pole locations relative to the measurement window, it is easy to justify that such an approximation will be valid for the positive sum of poles representing the polarizability of each state.  From theoretical calculations, the agreement between the Pad\`e approximant and the polarizability of each state is better than 0.2\% for $\lambda>630\,\mathrm{nm}$.  We stress that this agreement is a consequence of the pole locations and not dependent on specific contributions.  However, this is not necessarily true for the differential polarizability for which significant cancelation of terms can occur and even flip the relative sign of the quadratic and quartic terms.

Since the polarizability of each state can be well-represented by a SPM, it follows that a five parameter differential pole model (DPM) given by
\begin{equation}
\Delta\alpha_0(\omega)=c_0+c_f\frac{(\omega/\omega_f)^2}{1-(\omega/\omega_f)^2}-c_g\frac{(\omega/\omega_g)^2}{1-(\omega/\omega_g)^2},
\end{equation}
is a valid model.  However, there is insufficient data for this to be used.  Moreover, if data fits to a SPM, then there would be no ability to distinguish the $c_f$ and $c_g$ terms.  To illustrate that the DPM can significantly change an extrapolation, we can add some additional constraints.  We first note that the fit to the SPM gives a zero crossing near 680\,nm.  Since direct confirmation of this zero crossing has been presented at several conferences by the authors of \cite{huntemann2016single},  we constrain the zero crossing to this value.  For any given pair of poles $(\omega_g,\omega_f)$, we can then fit the DPM to the data in table~\ref{YbData} as a two parameter fit to four data points.

In Fig.~\ref{offset}(b) we plot the fit to a DPM in which the pole location for the ground-state is constrained to 337\,nm and for the upper state to 276\,nm, which are the approximate locations predicted by theory.  The fitted estimate for the dc value is $5.11(15)\,\mathrm{a.u}$., although we do not suggest this is a meaningful estimate.  The fitted curve almost exactly agrees with the value at $7.17\,\mathrm{\mu m}$ reported in \cite{baynham2018measurement}, which is indicated by the blue diamond.  Also, the value at $7.17\,\mathrm{\mu m}$ is within $0.06\%$ of the dc value due to a significant cancelation of the quadratic terms in the expansion for $\Delta\alpha_0(\omega)$.  We do not suggest this is the case in reality, only that this particular fit has this behaviour.

The dc estimate from the DPM is actually insensitive to the choice of pole positions.  Changing the pole positions by as much as $\pm10\,\mathrm{nm}$ changes the estimated value of $\Delta\alpha_0(0)$ by no more the 13\% of its estimated uncertainty.  Consequently, it is tempting to think the estimate is reasonable.  However, one must take into account that the model is based on the physics of the atom and the fitted parameters are not arbitrary.  The pole position of 337\,nm for the ground state is primarily determined by transitions to the $^2\mathrm{P}_{1/2}$ and $^2\mathrm{P}_{3/2}$ levels at 369.4\,nm and 328.9\,nm respectively, and $c_g$ by their corresponding line strengths.  These two transitions make up 72\% of the calculated dc polarizability of the ground state and the measured lifetime of $^2\mathrm{P}_{1/2}$ \cite{olmschenk2006precision} is within 6.5\% of the calculated value \cite{biemont1998lifetime}.  Similarly $\omega_f$ and $c_f$ are primarily determined by transitions at 275.0\,nm and 265.4\,nm with the next two largest contributions at 286.0\,nm and 267.3\,nm.  These four transitions make up 61\% of the calculated upper-state dc polarizability and measured lifetimes of the associated upper states have better than 4\% agreement with theory \cite{pinnington1994beam,biemont1998lifetime}.  However, the fitted values of $c_g$ and $c_f$ are roughly 3 times larger than the values estimated from theory.  Given the level of agreement between lifetime measurements reported in \cite{olmschenk2006precision,pinnington1994beam} and theory \cite{biemont1998lifetime}, such large values for $c_g$ and $c_f$ would suggest a substantial departure from theory and/or the results in \cite{olmschenk2006precision,pinnington1994beam} incorrect.  Reasonable changes to the pole positions ($\pm 10\,\mathrm{nm}$) do not change these conclusions.

Results using the projection method are given in table~\ref{YbProjector}.  The first set of values in the table include only the four measurements in table~\ref{YbData}.  The $k=2$ term gives an estimate that is consistent with offsetting the calculated $\Delta\alpha_0(\omega)$ by a constant, albeit with a larger uncertainty, which is a consequence of allowing variation in the curvature consistent with the measurements and the structure of the atom.  The residual correction term is on the order of the uncertainty $\|\mathbf{x}\|$, but it may be considered questionable to make a correction if there was no faith in the theory.  However, taking the uncertainty in this correction as the value itself is equivalent to assuming all contributions have a 43\% uncertainty with maximal correlation or more than 100\% uncertainty with no correlation.  All things considered  \cite{olmschenk2006precision,pinnington1994beam,biemont1998lifetime}, this seems fairly reasonable and gives an estimate $\Delta\alpha_0(0)=5.534(85)\,\mathrm{a.u.}$, which is actually more precise than the incorrectly determined uncertainty in \cite{huntemann2016single}.   This highlights the fact that the value is not a conventional extrapolation but a bound placed on the allowed variations that are consistent with the atomic structure and the available measurements.  Additional constraints on remaining variations are imposed through an assumed knowledge of $\mathbf{c}$.

The estimate of $\Delta\alpha_0(0)=5.534(85)\,\mathrm{a.u.}$ from the $k=2$ term is dominated by the measurements at 1310\,nm, which has the highest precision by a factor of 6, and that at 852\,nm, which is furthest from the 1310\,nm measurement.  The other two measurements have a relatively small weight.  This highlights the redundancy in the measurements insofar as extrapolation is concerned.  The measurement at 1554\,nm is too close to the more precise measurement at 1310\,nm to provide any further information.  Similarly the measurement at 1064\,nm provides no significant information on changes in $\Delta\alpha_0(\omega)$ not already captured by the measurement at 852\,nm.  Indeed, measurements at 1310\,nm, 1064\,nm and 852\,nm credibly fit to a straight line and all four data points provide limited constraint on possible curvature.  Consequently any multi-parameter modelling of the data is likely to lead to over-fitting, and this is fairly self-evident from the plots in Fig.~\ref{offset}.

The $k=3$ term eliminates any practical dependence on theory, but the added contribution is statistically meaningless indicating that the measurements are insufficient to provide an accurate, strictly measurement-based extrapolation.  With the dependence on $\mathbf{c}$ eliminated, the procedure then becomes a fit to a model which is at least consistent with the atomic structure of the atom.  Over the measurement window of interest, that model is something between a SPM or a general quartic polynomial restricted to even order terms.  Hence we obtain an estimate of $5.34(43)\,\mathrm{a.u.}$, which is consistent with the $k=2$ estimate and that from a SPM albeit with a larger uncertainty. 
\begin{table}
\caption{Extrapolations of $\Delta \alpha_0(\omega)$ to dc using the projection method on Yb$^+$ measurements listed in table~\ref{YbData}.  The second and third set of tabulated values includes a zero crossing at 680\,nm, and 635\,nm respectively.}
\begin{ruledtabular}
\begin{tabular}{c c c c c c}
$k$ & $\bar{\mathbf{m}}\cdot\mathbf{x}$ & $\|\mathbf{x}\|$ & $\mathbf{c}\cdot\bar{\mathbf{u}}$ & $\sigma_\mathrm{rms}$ & $\sigma_c$\\ [0.1pc]
\hline \\ [-0.6pc]
2 & 5.607 & 0.067 & -0.0533 & 0.0434 & 0.1215\\
3 & 5.347 & 0.430 & -0.0006 & 0.0012 & 0.0023 \\
\hline \\ [-0.6pc]
2 & 5.953 & 0.032 & -0.0892 & 0.0783 & 0.2201\\
3 & 5.170 & 0.136 & -0.0024 & 0.0039 & 0.0076 \\
\hline \\ [-0.6pc]
2 & 5.676 & 0.029 & -0.1042 & 0.0950 & 0.2672\\
3 & 5.536 & 0.115 & -0.0030 & 0.0048 & 0.0091 \\
\end{tabular}
\end{ruledtabular}
\label{YbProjector}
\end{table}

The second set of values in table~\ref{YbProjector} includes a measurement of $0.00(0.05)\,\mathrm{a.u.}$ at 680\,nm as a zero crossing determination.  In this case, the $k=3$ term is statistically meaningful and eliminates the theory contribution.  However the correction of -0.78(13) is substantially larger than the $k=2$ theory term would suggest being $\gtrsim3\sigma_c$.  Recalling that $\sigma_c$ accounts for a 100\% error in the theoretical estimates with maximal correlation, the degree to which theory would have to be corrected seems unlikely.  Even a $\pm10\,\mathrm{nm}$ change to the zero crossing does not change this observation.  This is consistent with the previously noted discrepancy found in $c_g$ and $c_f$ when fitting with a DPM.  The value of $5.17(12)\,\mathrm{a.u.}$ is also consistent with that determined from the DPM.

A discrepancy with theory can equally result from an inconsistency in the measurements.  To illustrate, consider a zero crossing measurement giving $0.00(0.05)\,\mathrm{a.u.}$ at 635\,nm, which results in the third set of tabulated values in table~\ref{YbProjector}.  The $k=3$ correction eliminates the contribution from theory and is consistent with the $k=2$ theory contribution.  However, it lacks statistical significance again indicating an insufficient measurement window.  For comparison a fit using the DPM is shown in Fig.~\ref{offset}(c) where we have used the same pole locations as for the fit in Fig.~\ref{offset}(b).  The fit gives an estimate of $5.52(0.12)$ consistent with the projection method and the coefficients $c_g$ and $c_f$ from the DPM fit agree with the theory values to within a few percent.

It is illuminating to consider what happens as the zero crossing is moved from shorter to longer wavelengths.  Constrained by the data in table~\ref{YbData}, a zero crossing at increasing wavelengths forces an increasing curvature in $\Delta\alpha_0(\omega)$.  Given the pole positions of the basis functions relative to the measurement window of interest, this requires an increasingly large and eventually unrealistic deviation from theoretical estimates of $\mathbf{c}$.  This is supported from the considerations of either the projection method or a DPM.  Additionally, as the zero crossing is moved from shorter to longer wavelengths, the $k=3$ term has increasing statistical significance.  This reflects the fact that measurements increasingly allow a better determination of the curvature and hence frequency dependence of $\Delta\alpha_0(\omega)$.

\section{Discussion}
We have presented a numerical method for extrapolating polarisability measurements to dc as done in the assessment of blackbody radiation shifts for ion-based clocks.  Our method explicitly accounts for the frequency dependence allowed by the available atomic transitions without introducing an ad hoc modelling function.  Although proper interpretation requires \emph{a priori} atomic structure calculations, it is not dependent on the accuracy of those calculations provided there are sufficient measurements spanning a suitable measurement window as needed for any extrapolation.  Inclusion of the available calculations also provides indicators of inconsistencies between theory and experiment, or inadequacies of the experimental data.  

The method can be viewed as an extension and correction to that used in \cite{rosenband2006blackbody,brewer2019quantum} to more than one measurement.  In the case of Al$^+$, an adequate, fully experimental assessment of $\Delta\alpha_0(0)$ would require two moderately accurate measurements at a flexible selection of wavelengths.  As shown here, the projection method is consistent with that expected from a quadratic fit.  However, as the projection method properly accounts for the frequency dependence of the atomic transitions, it allows the measurement window to extend to a larger range of wavelengths.  For example, a measurement at the 397\,nm cooling transition of the Ca$^+$ logic ion in addition to the measurement at 987\,nm \cite{brewer2019quantum}, would have a 2.2\% bias in the extrapolated value compared to a 7.6\% bias from a quadratic fit.  

As an application to a single measurement, the method is essentially a correction to that given in \cite{rosenband2006blackbody,brewer2019quantum}.  Although the difference between the two methods is small in the case of Al$^+$, this will not be true in general.  More importantly, the application to a single measurement is a theoretical correction, which requires consistency between experiment and theory.  This can come from external evidence but, at a minimum, the measured value should be in reasonable agreement with the theory used.  This is not so easy to quantify, but if the correction was significantly larger than the measurement precision, one should probably consider another measurement at a well separated wavelength.  We would also suggest that an assessment using theoretical calculations should not so readily disregard possible correlation in calculated matrix elements.

Problems with the analysis in \cite{huntemann2016single} aside, application of the projection method to Yb$^+$ demonstrates some interesting features of the method.  It highlights that the projection method is not a conventional extrapolation.  It bounds maximally allowed variations through constraints imposed by the available measurements and knowledge of the contributions $\mathbf{c}$ to the polarizability.  Importantly, it removes any possible ambiguity that might arise in choosing a fitting function for making an estimate for points outside a given measurement window.  

In \cite{huntemann2016single}, use of NIR radiation was motivated by noting that all transitions contributing to $\alpha_0(\omega)$ are below $380\,\mathrm{nm}$ and cited the approach used for Al$^+$ \cite{rosenband2006blackbody}.  However, no extrapolation was carried out \cite{rosenband2006blackbody}.  Rather a single measurement was made and a theoretical argument given to justify a small correction to the measurement.  The projection method provides the correct mathematical framework for that procedure that can be applied to any number of measurements.  Applied to the measurements in \cite{huntemann2016single}, the procedure yields $5.534(85)\,\mathrm{a.u.}$ when allowing a fairly substantial uncertainty in the calculations, but this disregards a zero crossing determination, which introduces an inconsistency depending on its location.

For Yb$^+$, location of the zero crossing has a significant influence on the inferred value of $\Delta\alpha_0(0)$.  Both the projection method and considerations using a DPM model strongly suggest that a zero crossing at 680\,nm is unlikely, which is also supported by the SPM fit itself.  A SPM is a valid model up to a $4^\mathrm{th}$ order expansion provided there is no inflection point i.e. a relative sign change between the quadratic and quartic terms.  Thus, it can only faithfully represent $\Delta\alpha_0(\omega)$ over a limited range.  Since the fitted pole position is at 540\,nm and all contributing poles are below 380\,nm, the range of validity would have to be limited to the $4^\mathrm{th}$ order expansion.  This is marginally true at the 852\,nm edge of the measurement window used in \cite{huntemann2016single}, which allows a reasonable extrapolation to dc, but it is certainly not true at the zero crossing of 680\,nm. The rapidly changing SPM in the vicinity of 680\,nm can only be matched by substantial changes to the line strengths of the real transitions below 380\,nm and those changes have to be consistent with measurements reported in \cite{huntemann2016single,olmschenk2006precision,pinnington1994beam}, which places strict bounds on the dominant contributions to $\Delta\alpha_0(\omega)$ over the region of interest. 

A question arises as to the credibility of a Yb$^+$ E3 clock assessment.  Clearly the reported uncertainty in \cite{huntemann2016single} is substantially incorrect.  Correct error analysis would increase the uncertainty to more than $10^{-17}$.  The projection method reported here would credibly restore the accuracy claim with the estimate of $\Delta\alpha_0(0)=5.534(85)\,\mathrm{a.u.}$, but that estimate, or any other estimate for that matter, is completely undermined by the zero crossing at 680\,nm.  Consequently, any credible assessment of the Yb$^+$ E3 clock transition would have to include a report of the zero crossing and address the inconsistencies with existing data that it may represent.

\begin{acknowledgements}
This research is supported by the Agency for Science, Technology and Research (A*STAR) under Project No. C210917001; the National Research Foundation (NRF), Singapore, under its Quantum Engineering Programme (QEP-P5); the National Research Foundation, Singapore and A*STAR under its Quantum Engineering Programme (NRF2021-QEP2-01-P03) and its CQT Bridging Grant; and the Ministry of Education, Singapore under its Academic Research Fund Tier 2 (MOE-T2EP50120-0014).
\end{acknowledgements} 
\appendix
\section{SPM model for each clock state}
Parameters for the Pad\`e approximant for either clock state are given by,
\begin{equation*}
c_0=\alpha_0(0), \quad c_1=\frac{6\left(\alpha^{(2)}_0(0)\right)^2}{\alpha^{(4)}_0(0)},
\end{equation*}
and
\begin{equation*}
\omega_0=\sqrt{\frac{12\alpha^{(2)}_0(0)}{\alpha^{(4)}_0(0)}},
\end{equation*}
where $\alpha_0(\omega)$ is the polarizability for the state of interest and $\alpha^{(n)}_0(0)$ denotes the $n^\mathrm{th}$ derivative.  From these expressions, the DPM determined from theory has $c_g\approx50$, $c_f \approx 58$ and corresponding pole positions at $\lambda_g=337\,\mathrm{nm}$ and  $\lambda_f=276\,\mathrm{nm}$.  

To investigate variations in the parameters due to changes in theory, we take matrix elements to be normally distributed random variables with means determined from \cite{biemont1998lifetime}.  Standard deviations for the 265-, 267-, 275-, 286-, and 369\,nm matrix elements are taken to be consistent with the uncertainty in the measured lifetime of their associated upper states \cite{olmschenk2006precision,pinnington1994beam}.  Standard deviations for all other matrix elements are taken to be 50\% with the exception of the 329\,nm transition, which is taken to be 5\%, as we expect the ratio of matrix elements for the 329 and 369 transitions to be more accurately calculated.  Calculated parameters under these variations are reasonably well described by normal distributions.  For the ground state we obtain $c_g=53(5)$ with a pole position 335(4)\,nm.  For the $F$ state, we obtain $c_f=63(5)$ with a pole position of 276(7)\,nm.  Mean values are slightly shifted from the values obtained from fixed values as the square of a normal distribution is governed by a non-central chi-square distribution, which shifts the mean.
\bibliography{Polarizability}

\begin{thebibliography}{11}%
\makeatletter
\providecommand \@ifxundefined [1]{%
 \@ifx{#1\undefined}
}%
\providecommand \@ifnum [1]{%
 \ifnum #1\expandafter \@firstoftwo
 \else \expandafter \@secondoftwo
 \fi
}%
\providecommand \@ifx [1]{%
 \ifx #1\expandafter \@firstoftwo
 \else \expandafter \@secondoftwo
 \fi
}%
\providecommand \natexlab [1]{#1}%
\providecommand \enquote  [1]{``#1''}%
\providecommand \bibnamefont  [1]{#1}%
\providecommand \bibfnamefont [1]{#1}%
\providecommand \citenamefont [1]{#1}%
\providecommand \href@noop [0]{\@secondoftwo}%
\providecommand \href [0]{\begingroup \@sanitize@url \@href}%
\providecommand \@href[1]{\@@startlink{#1}\@@href}%
\providecommand \@@href[1]{\endgroup#1\@@endlink}%
\providecommand \@sanitize@url [0]{\catcode `\\12\catcode `\$12\catcode
  `\&12\catcode `\#12\catcode `\^12\catcode `\_12\catcode `\%12\relax}%
\providecommand \@@startlink[1]{}%
\providecommand \@@endlink[0]{}%
\providecommand \url  [0]{\begingroup\@sanitize@url \@url }%
\providecommand \@url [1]{\endgroup\@href {#1}{\urlprefix }}%
\providecommand \urlprefix  [0]{URL }%
\providecommand \Eprint [0]{\href }%
\providecommand \doibase [0]{http://dx.doi.org/}%
\providecommand \selectlanguage [0]{\@gobble}%
\providecommand \bibinfo  [0]{\@secondoftwo}%
\providecommand \bibfield  [0]{\@secondoftwo}%
\providecommand \translation [1]{[#1]}%
\providecommand \BibitemOpen [0]{}%
\providecommand \bibitemStop [0]{}%
\providecommand \bibitemNoStop [0]{.\EOS\space}%
\providecommand \EOS [0]{\spacefactor3000\relax}%
\providecommand \BibitemShut  [1]{\csname bibitem#1\endcsname}%
\let\auto@bib@innerbib\@empty
\bibitem [{\citenamefont {Arnold}\ \emph {et~al.}(2019)\citenamefont {Arnold},
  \citenamefont {Kaewuam}, \citenamefont {Tan}, \citenamefont {Porsev},
  \citenamefont {Safronova},\ and\ \citenamefont
  {Barrett}}]{arnold2019dynamic}%
  \BibitemOpen
  \bibfield  {author} {\bibinfo {author} {\bibfnamefont {K.}~\bibnamefont
  {Arnold}}, \bibinfo {author} {\bibfnamefont {R.}~\bibnamefont {Kaewuam}},
  \bibinfo {author} {\bibfnamefont {T.}~\bibnamefont {Tan}}, \bibinfo {author}
  {\bibfnamefont {S.}~\bibnamefont {Porsev}}, \bibinfo {author} {\bibfnamefont
  {M.}~\bibnamefont {Safronova}}, \ and\ \bibinfo {author} {\bibfnamefont
  {M.}~\bibnamefont {Barrett}},\ }\href@noop {} {\bibfield  {journal} {\bibinfo
   {journal} {Phys. Rev. A}\ }\textbf {\bibinfo {volume} {99}},\ \bibinfo
  {pages} {012510} (\bibinfo {year} {2019})}\BibitemShut {NoStop}%
\bibitem [{\citenamefont {Huntemann}\ \emph {et~al.}(2016)\citenamefont
  {Huntemann}, \citenamefont {Sanner}, \citenamefont {Lipphardt}, \citenamefont
  {Tamm},\ and\ \citenamefont {Peik}}]{huntemann2016single}%
  \BibitemOpen
  \bibfield  {author} {\bibinfo {author} {\bibfnamefont {N.}~\bibnamefont
  {Huntemann}}, \bibinfo {author} {\bibfnamefont {C.}~\bibnamefont {Sanner}},
  \bibinfo {author} {\bibfnamefont {B.}~\bibnamefont {Lipphardt}}, \bibinfo
  {author} {\bibfnamefont {C.}~\bibnamefont {Tamm}}, \ and\ \bibinfo {author}
  {\bibfnamefont {E.}~\bibnamefont {Peik}},\ }\href@noop {} {\bibfield
  {journal} {\bibinfo  {journal} {Phys. Rev. Lett.}\ }\textbf {\bibinfo
  {volume} {116}},\ \bibinfo {pages} {063001} (\bibinfo {year}
  {2016})}\BibitemShut {NoStop}%
\bibitem [{\citenamefont {Rosenband}\ \emph {et~al.}(2006)\citenamefont
  {Rosenband}, \citenamefont {Itano}, \citenamefont {Schmidt}, \citenamefont
  {Hume}, \citenamefont {Koelemeij}, \citenamefont {Bergquist},\ and\
  \citenamefont {Wineland}}]{rosenband2006blackbody}%
  \BibitemOpen
  \bibfield  {author} {\bibinfo {author} {\bibfnamefont {T.}~\bibnamefont
  {Rosenband}}, \bibinfo {author} {\bibfnamefont {W.~M.}\ \bibnamefont
  {Itano}}, \bibinfo {author} {\bibfnamefont {P.}~\bibnamefont {Schmidt}},
  \bibinfo {author} {\bibfnamefont {D.}~\bibnamefont {Hume}}, \bibinfo {author}
  {\bibfnamefont {J.}~\bibnamefont {Koelemeij}}, \bibinfo {author}
  {\bibfnamefont {J.~C.}\ \bibnamefont {Bergquist}}, \ and\ \bibinfo {author}
  {\bibfnamefont {D.~J.}\ \bibnamefont {Wineland}},\ }in\ \href@noop {} {\emph
  {\bibinfo {booktitle} {Proceedings of the 20th European Frequency and Time
  Forum}}}\ (\bibinfo {organization} {IEEE},\ \bibinfo {year} {2006})\ pp.\
  \bibinfo {pages} {289--292}\BibitemShut {NoStop}%
\bibitem [{\citenamefont {Safronova}\ \emph {et~al.}(2012)\citenamefont
  {Safronova}, \citenamefont {Kozlov},\ and\ \citenamefont
  {Clark}}]{safronova2012blackbody}%
  \BibitemOpen
  \bibfield  {author} {\bibinfo {author} {\bibfnamefont {M.~S.}\ \bibnamefont
  {Safronova}}, \bibinfo {author} {\bibfnamefont {M.~G.}\ \bibnamefont
  {Kozlov}}, \ and\ \bibinfo {author} {\bibfnamefont {C.~W.}\ \bibnamefont
  {Clark}},\ }\href@noop {} {\bibfield  {journal} {\bibinfo  {journal} {IEEE
  transactions on ultrasonics, ferroelectrics, and frequency control}\ }\textbf
  {\bibinfo {volume} {59}},\ \bibinfo {pages} {439} (\bibinfo {year}
  {2012})}\BibitemShut {NoStop}%
\bibitem [{\citenamefont {Brewer}\ \emph {et~al.}(2019)\citenamefont {Brewer},
  \citenamefont {Chen}, \citenamefont {Hankin}, \citenamefont {Clements},
  \citenamefont {Chou}, \citenamefont {Wineland}, \citenamefont {Hume},\ and\
  \citenamefont {Leibrandt}}]{brewer2019quantum}%
  \BibitemOpen
  \bibfield  {author} {\bibinfo {author} {\bibfnamefont {S.}~\bibnamefont
  {Brewer}}, \bibinfo {author} {\bibfnamefont {J.-S.}\ \bibnamefont {Chen}},
  \bibinfo {author} {\bibfnamefont {A.}~\bibnamefont {Hankin}}, \bibinfo
  {author} {\bibfnamefont {E.}~\bibnamefont {Clements}}, \bibinfo {author}
  {\bibfnamefont {C.}~\bibnamefont {Chou}}, \bibinfo {author} {\bibfnamefont
  {D.}~\bibnamefont {Wineland}}, \bibinfo {author} {\bibfnamefont
  {D.}~\bibnamefont {Hume}}, \ and\ \bibinfo {author} {\bibfnamefont
  {D.}~\bibnamefont {Leibrandt}},\ }\href@noop {} {\bibfield  {journal}
  {\bibinfo  {journal} {Phys. Rev. Lett.}\ }\textbf {\bibinfo {volume} {123}},\
  \bibinfo {pages} {033201} (\bibinfo {year} {2019})}\BibitemShut {NoStop}%
\bibitem [{\citenamefont {Arnold}\ \emph {et~al.}(2018)\citenamefont {Arnold},
  \citenamefont {Kaewuam}, \citenamefont {Roy}, \citenamefont {Tan},\ and\
  \citenamefont {Barrett}}]{arnold2018blackbody}%
  \BibitemOpen
  \bibfield  {author} {\bibinfo {author} {\bibfnamefont {K.}~\bibnamefont
  {Arnold}}, \bibinfo {author} {\bibfnamefont {R.}~\bibnamefont {Kaewuam}},
  \bibinfo {author} {\bibfnamefont {A.}~\bibnamefont {Roy}}, \bibinfo {author}
  {\bibfnamefont {T.}~\bibnamefont {Tan}}, \ and\ \bibinfo {author}
  {\bibfnamefont {M.}~\bibnamefont {Barrett}},\ }\href@noop {} {\bibfield
  {journal} {\bibinfo  {journal} {Nat. Comm.}\ }\textbf {\bibinfo {volume}
  {9}},\ \bibinfo {pages} {1650} (\bibinfo {year} {2018})}\BibitemShut
  {NoStop}%
\bibitem [{\citenamefont {Paez}\ \emph {et~al.}(2016)\citenamefont {Paez},
  \citenamefont {Arnold}, \citenamefont {Hajiyev}, \citenamefont {Porsev},
  \citenamefont {Dzuba}, \citenamefont {Safronova}, \citenamefont {Safronova},\
  and\ \citenamefont {Barrett}}]{paez2016atomic}%
  \BibitemOpen
  \bibfield  {author} {\bibinfo {author} {\bibfnamefont {E.}~\bibnamefont
  {Paez}}, \bibinfo {author} {\bibfnamefont {K.}~\bibnamefont {Arnold}},
  \bibinfo {author} {\bibfnamefont {E.}~\bibnamefont {Hajiyev}}, \bibinfo
  {author} {\bibfnamefont {S.}~\bibnamefont {Porsev}}, \bibinfo {author}
  {\bibfnamefont {V.}~\bibnamefont {Dzuba}}, \bibinfo {author} {\bibfnamefont
  {U.}~\bibnamefont {Safronova}}, \bibinfo {author} {\bibfnamefont
  {M.}~\bibnamefont {Safronova}}, \ and\ \bibinfo {author} {\bibfnamefont
  {M.}~\bibnamefont {Barrett}},\ }\href@noop {} {\bibfield  {journal} {\bibinfo
   {journal} {Physical Review A}\ }\textbf {\bibinfo {volume} {93}},\ \bibinfo
  {pages} {042112} (\bibinfo {year} {2016})}\BibitemShut {NoStop}%
\bibitem [{\citenamefont {Bi{\'e}mont}\ \emph {et~al.}(1998)\citenamefont
  {Bi{\'e}mont}, \citenamefont {Dutrieux}, \citenamefont {Martin},\ and\
  \citenamefont {Quinet}}]{biemont1998lifetime}%
  \BibitemOpen
  \bibfield  {author} {\bibinfo {author} {\bibfnamefont {E.}~\bibnamefont
  {Bi{\'e}mont}}, \bibinfo {author} {\bibfnamefont {J.}~\bibnamefont
  {Dutrieux}}, \bibinfo {author} {\bibfnamefont {I.}~\bibnamefont {Martin}}, \
  and\ \bibinfo {author} {\bibfnamefont {P.}~\bibnamefont {Quinet}},\
  }\href@noop {} {\bibfield  {journal} {\bibinfo  {journal} {Journal of Physics
  B: Atomic, Molecular and Optical Physics}\ }\textbf {\bibinfo {volume}
  {31}},\ \bibinfo {pages} {3321} (\bibinfo {year} {1998})}\BibitemShut
  {NoStop}%
\bibitem [{\citenamefont {Baynham}\ \emph {et~al.}(2018)\citenamefont
  {Baynham}, \citenamefont {Curtis}, \citenamefont {Godun}, \citenamefont
  {Jones}, \citenamefont {Nisbet-Jones}, \citenamefont {Baird}, \citenamefont
  {Bongs}, \citenamefont {Gill}, \citenamefont {Fordell}, \citenamefont {Hieta}
  \emph {et~al.}}]{baynham2018measurement}%
  \BibitemOpen
  \bibfield  {author} {\bibinfo {author} {\bibfnamefont {C.}~\bibnamefont
  {Baynham}}, \bibinfo {author} {\bibfnamefont {E.}~\bibnamefont {Curtis}},
  \bibinfo {author} {\bibfnamefont {R.}~\bibnamefont {Godun}}, \bibinfo
  {author} {\bibfnamefont {J.}~\bibnamefont {Jones}}, \bibinfo {author}
  {\bibfnamefont {P.}~\bibnamefont {Nisbet-Jones}}, \bibinfo {author}
  {\bibfnamefont {P.}~\bibnamefont {Baird}}, \bibinfo {author} {\bibfnamefont
  {K.}~\bibnamefont {Bongs}}, \bibinfo {author} {\bibfnamefont
  {P.}~\bibnamefont {Gill}}, \bibinfo {author} {\bibfnamefont {T.}~\bibnamefont
  {Fordell}}, \bibinfo {author} {\bibfnamefont {T.}~\bibnamefont {Hieta}},
  \emph {et~al.},\ }\href@noop {} {\bibfield  {journal} {\bibinfo  {journal}
  {arXiv preprint arXiv:1801.10134}\ } (\bibinfo {year} {2018})}\BibitemShut
  {NoStop}%
\bibitem [{\citenamefont {Olmschenk}\ \emph {et~al.}(2006)\citenamefont
  {Olmschenk}, \citenamefont {Hayes}, \citenamefont {Matsukevich},
  \citenamefont {Maunz}, \citenamefont {Moehring}, \citenamefont {Younge},\
  and\ \citenamefont {Monroe}}]{olmschenk2006precision}%
  \BibitemOpen
  \bibfield  {author} {\bibinfo {author} {\bibfnamefont {S.}~\bibnamefont
  {Olmschenk}}, \bibinfo {author} {\bibfnamefont {D.}~\bibnamefont {Hayes}},
  \bibinfo {author} {\bibfnamefont {D.}~\bibnamefont {Matsukevich}}, \bibinfo
  {author} {\bibfnamefont {P.}~\bibnamefont {Maunz}}, \bibinfo {author}
  {\bibfnamefont {D.}~\bibnamefont {Moehring}}, \bibinfo {author}
  {\bibfnamefont {K.}~\bibnamefont {Younge}}, \ and\ \bibinfo {author}
  {\bibfnamefont {C.}~\bibnamefont {Monroe}},\ }\href@noop {} {\bibfield
  {journal} {\bibinfo  {journal} {Phys. Rev. A}\ }\textbf {\bibinfo {volume}
  {73}},\ \bibinfo {pages} {023413} (\bibinfo {year} {2006})}\BibitemShut
  {NoStop}%
\bibitem [{\citenamefont {Pinnington}\ \emph {et~al.}(1994)\citenamefont
  {Pinnington}, \citenamefont {Berends},\ and\ \citenamefont
  {Ji}}]{pinnington1994beam}%
  \BibitemOpen
  \bibfield  {author} {\bibinfo {author} {\bibfnamefont {E.}~\bibnamefont
  {Pinnington}}, \bibinfo {author} {\bibfnamefont {R.}~\bibnamefont {Berends}},
  \ and\ \bibinfo {author} {\bibfnamefont {Q.}~\bibnamefont {Ji}},\ }\href@noop
  {} {\bibfield  {journal} {\bibinfo  {journal} {Physical Review A}\ }\textbf
  {\bibinfo {volume} {50}},\ \bibinfo {pages} {2758} (\bibinfo {year}
  {1994})}\BibitemShut {NoStop}%
\end{thebibliography}%
\end{document}